\documentclass{mn2e}
\usepackage{color}
\usepackage{natbib}
\usepackage{amsfonts}
\usepackage{amsmath}
\usepackage{amssymb}
\usepackage{graphicx}

\newcommand{\p}{\ensuremath{\partial}}
\newcommand{\del}{\ensuremath{\delta}}
\newcommand{\Del}{\ensuremath{\Delta}}
\newcommand{\lam}{\ensuremath{\lambda}}
\newcommand{\gam}{\ensuremath{\gamma}}
\newcommand{\Gam}{\ensuremath{\Gamma}}
\newcommand{\sig}{\ensuremath{\sigma}}

\newcommand{\epc}{\ensuremath{\epsilon_{\times}}}
\newcommand{\Sc}{\ensuremath{S_{\times}}}
\newcommand{\So}{\ensuremath{S_{0}}}
\newcommand{\delc}{\ensuremath{\delta_{\rm c}}}
\newcommand{\delo}{\ensuremath{\delta_{0}}}
\newcommand{\rhoh}{\ensuremath{\rho_{\rm h}}}
\newcommand{\xtil}{\ensuremath{\tilde x}}

\newcommand{\Msun}{\ensuremath{M_{\odot}}}

\newcommand{\avg}[1]{\ensuremath{\left\langle \,#1\, \right\rangle}}
\newcommand{\etal}{et al.}

\newcommand{\der}{\ensuremath{{\rm d}}}
\newcommand{\dir}{\ensuremath{\delta_{\rm D}}}

\newcommand{\cb}{\ensuremath{\mathbf{c}}}

\newcommand{\eqn}[1]{equation~\eqref{#1}}
\newcommand{\eqns}[1]{equations~\eqref{#1}}
\newcommand{\fig}[1]{Figure~\ref{#1}}

\newcommand{\ph}[1]{\phantom{#1}}

\newcommand{\be}{\begin{equation}}
\newcommand{\ee}{\end{equation}}
\newcommand{\Cal}[1]{\ensuremath{\mathcal{#1}}}

\title[Excursion set peaks]
      {Excursion set peaks:  a self-consistent model of dark halo abundances and clustering} 

\author[A. Paranjape, R.K. Sheth \& V. Desjacques]
{Aseem Paranjape$^{1,2}$\thanks{E-mail: aseemp@phys.ethz.ch}, Ravi K. Sheth$^{1,3}$\thanks{E-mail: sheth@ictp.it} \& 
Vincent Desjacques$^{4}$\thanks{E-mail: Vincent.Desjacques@unige.ch}\\  
 $^1$ The Abdus Salam International Center for Theoretical Physics,
      Strada Costiera, 11, Trieste 34151, Italy\\
 $^2$ ETH Z\"urich, Instt. for Astronomy, Wolfgang-Pauli-Strasse 27, CH-8093 Z\"urich, Switzerland\\
 $^3$ Center for Particle Cosmology, University of Pennsylvania, 
      209 S. 33rd St., Philadelphia, PA 19104, USA\\
 $^4$ D\'epartement de Physique Th\'eorique and 
      Center for Astroparticle Physics (CAP), Universit\'e de Gen\`eve, \\ 
      24 quai Ernest Ansermet, CH-1211 Gen\`eve, Switzerland}
\begin{document}
\pagerange{\pageref{firstpage}--\pageref{lastpage}}

\maketitle 

\label{firstpage}

\begin{abstract}
\noindent 
We describe how to extend the excursion set peaks framework so that its predictions of dark halo abundances and clustering can be compared directly with simulations.  These extensions include: a halo mass definition which uses the TopHat filter in real space; the mean dependence of the critical density for collapse \delc\ on halo mass $m$; and the scatter around this mean value.  All three of these are motivated by the physics of triaxial rather than spherical collapse.  A comparison of the resulting mass function with $N$-body results shows that, if one uses $\delc(m)$ and its scatter as determined from simulations, then all three are necessary ingredients for obtaining $\sim 10\%$ accuracy.  E.g., assuming a constant value of $\delc$ with no scatter, as motivated by the physics of spherical collapse, leads to many more massive halos than seen in simulations.  
The same model is also in excellent agreement with $N$-body results for the linear halo bias, especially at the high mass end where the traditional peak-background split argument applied to the mass function fit is known to underpredict the measured bias by $\sim10\%$.  In the excursion set language, our model is about walks centered on special positions (peaks) in the initial conditions -- we discuss what it implies for the usual calculation in which all walks contribute to the statistics.  
\end{abstract}


\section{Introduction}
\label{intro}

The evolution of the abundance of virialized clusters is a powerful probe of the combined effects of the nature of the initial conditions (Gaussian or not?), the nature of gravity (general relativity or not?), and the expansion history of the Universe (cosmological constant or more complex?).  This has motivated studies to understand and parametrize this dependence, so that datasets can more precisely constrain the physical models.  

There are three widespread approaches to modeling cluster abundances.  The most accurate of these is the brute force numerical simulation method; one simply asserts that the dark matter halos which form in these simulations represent clusters in the Universe. However, this method is computationally expensive owing to the nonlinear nature of gravitational structure formation.  
 Furthermore, while simulations can provide insights into structure formation, they cannot replace a deep understanding of the physical mechanisms.
The other two are simpler analytic models which consider the statistics of the initial fluctuation field to model halo formation.  One is based on peaks, and attempts to model the comoving number density ${\rm d}n/{\rm d}m$ of objects of a given mass $m$ \citep{bbks86, aj90, m+98, h01}.  The other, known as the excursion set approach, seeks to count the mass fraction $f(m)$ that is bound-up in objects of a given mass, and, from this, to infer the abundances of objects themselves \citep{ps74, ph90, bcek91, st02, mr10a, ms12}.  The two are related by the simple fact that mass weighting each halo must yield the mass fraction in halos, so 
\be
 \int_M^\infty {\der m}\,\frac{\der n(m)}{\der m}\,m 
 = \bar\rho \int_M^\infty f(m)\,{\der m}.
 \label{nmfm}
\ee
However, this subtle difference has led to rather different formulations of the problem.  Only recently have the two approaches been written in the same formalism \citep{ps12}, allowing a direct comparison of how they differ.  Whereas all previous excursion set calculations of $f$ were based on the statistics of {\em all} positions in the initial density fluctuation field, \cite{ps12} showed that the Excursion Set Peaks (ESP)  calculation of $f$ explicitly uses the statistics of a small subset of positions in the initial field (those associated with peaks).

Working with a special subset of positions is motivated by measurements extracted from simulations showing that the excursion set prediction for the mass of the object in which a randomly chosen particle ends up is almost always incorrect \citep{w96}, whereas the prediction for the subset of particles which are at the center of mass of the proto-halo is almost always quite accurate \citep*{smt01}.  Moreover, for this special subset of (center-of-mass) particles, the predicted mass depends on whether one assumes a spherical or a triaxial collapse -- and the prediction is more accurate if one uses the more physically appropriate (though by no means perfect!) triaxial collapse model. In other words, the physics and the statistics are simpler, and tell a consistent story only if one works directly with the special subset of positions around which collapse occurs.  More recent work has shown that proto-halo positions do indeed coincide with peaks in the initial field \citep{lp11}, and their initial velocities are biased with respect to that of the dark matter \citep*{elp12}, as is expected for peaks \citep{ds10}.  This provides additional motivation for the ESP approach.  

However, this approach raises the following technical problem.  The peaks approach uses the statistics of the smoothed density field, as well as its first and second derivatives.  As a result, the rms values of these quantities play an important role.  In the theory, these rms values are related to integrals over the initial power spectrum.  However, they depend on the form of the smoothing filter, and, for the TopHat smoothing filter which is expected to be most directly related to the physics of gravitational collapse, some of these integrals do not converge.  For Gaussian filters there is no such problem, and \cite{ps12} showed that, when expressed in suitably scaled (and natural) units $\nu$, the ESP prediction for halo abundances is in excellent agreement with simulations, when the measured halo abundances are also expressed in scaled units. Namely, if one sets 
\be
 \frac{m}{\bar\rho}\frac{\der n(m)}{\der \ln m}
 = \nu\, f(\nu)\,\left|\frac{\der\ln\nu}{\der\ln m}\right|, 
 \label{excsetansatz}
\ee
then the ESP $\nu f(\nu)$ is in good agreement with the distribution of $\ln\nu$ measured in simulations.  However, this agreement is only a partial success because the relation between halo mass and $\nu$ depends on the smoothing filter, and the ESP model employed a Gaussian filter whereas simulations use a TopHat.  Therefore, when expressed as a function of mass, the Gaussian smoothed ESP prediction is not a good description of ${\rm d}n/{\rm d}\ln m$ in simulations (because the Jacobian $d\ln\nu/d\ln m$ depends on the filtering kernel).  Moreover, the Gaussian ESP prediction assumed that halos form through a spherical collapse, even though this is an incorrect description of the physics of halo formation \citep{smt01}.  In fact, there is substantial scatter even in the triaxial collapse model \citep{smt01,rks09,dwbs08,rktz09,lp11}, so the latter also provides an incomplete description of collapse. There has been some discussion of how to include stochasticity into the predictions 
\citep*{bm96,cl01,st02,vd08,mr10b,ca11,pls12,scs12},
but none of these self-consistently combines the model for the stochasticity with the fact that estimates of this stochasticity, from simulations or from theory, are for the special subset of positions around which collapse occurs.

Therefore, the main goals of the present work are to include TopHat smoothing in the ESP formalism and incorporate the effects of more complicated collapse models which are seen in simulations.  All this is the subject of Section~\ref{mf-uncond}.  Some testable predictions which elucidate the relation between the ESP approach and the usual excursion set analysis of all (rather than special) positions in space are discussed in Section~\ref{allwalks}.  Section~\ref{bias} presents implications for the spatial distribution of these objects -- the ESP predictions for Lagrangian halo bias -- which can be used to further test the model.  A final section summarizes our results. We will show numerical results at redshift $z=0$ for two $\Lambda$CDM cosmologies: one with $(\Omega_m,\Omega_\Lambda,\sig_8,n_s,h) = (0.3,0.7,0.9,1.0,0.7)$ which we denote WMAP1 and the other with $(0.25,0.75,0.8,1.0,0.7)$ which we denote WMAP3.

\section{The unconditional mass function}
\label{mf-uncond}

\subsection{Notation}
\label{notation}
We will designate Gaussian filtered quantities with the subscript $G$. By contrast, TopHat filtered objects will have no subscript.  The Fourier transforms of the TopHat and Gaussian filters are $W(kR)=(3/kR)j_1(kR)$ and $W_{\rm G}(kR_{\rm G}) = {\rm e}^{-k^2R_{\rm G}^2/2}$, respectively, with $j_1(y)$ a spherical Bessel function. 

We will frequently need the following integrals over the linearly extrapolated dimensionless matter power spectrum $\Del^2(k)\equiv k^3P(k)/2\pi^2$:
\vskip 0.05in
\noindent
{\rm\bf TopHat-filtered variance:}
\begin{align}
 s \equiv \sig_0^2 &\equiv \avg{\del^2} = \int \der\ln k\, \Del^2(k)W(kR)^2\,.
\label{sig0}
\end{align}
We will exclusively use the notation $\nu\equiv\delc/\sig_0$ where $\delc=1.686$ is the usual spherical collapse threshold.
\vskip 0.05in
\noindent
{\rm\bf Gaussian-filtered spatial moments:}
\begin{align}
\sig_{j{\rm G}}^2 &\equiv \int \der\ln k\, \Del^2(k)\,k^{2j}W_{\rm G}(kR_{\rm G})^2 ~~,~ j\geq1\,,
\label{sigjG}
\end{align}
so that $\sig_{1{\rm G}}^2 = \avg{(\nabla\del_{\rm G})^2}$ and $\sig_{2{\rm G}}^2 = \avg{(\nabla^2\del_{\rm G})^2}$.
\vskip 0.05in
\noindent
{\rm\bf Mixed moment:}
\begin{align}
\sig_{1{\rm m}}^2 &\equiv \avg{-\del\,\nabla^2\del_{\rm G}} \notag\\
&= \int \der\ln k\, \Del^2(k)\,k^2 W_{\rm G}(kR_{\rm G})W(kR)\,.
\label{sig1m}
\end{align}
\vskip 0.05in
\noindent
{\rm\bf Characteristic volume:}
\be
R_{\ast} \equiv \sqrt{3}\sig_{1{\rm G}}/\sig_{2{\rm G}} \qquad;\qquad  
V_{\ast} = (2\pi)^{3/2} R_{\ast}^3\,.
\label{VstRstgam}
\ee
These are the same as defined by \citet{bbks86}.
\vskip 0.05in
\noindent
{\rm\bf Spectral parameter:}
\begin{align}
\gam \equiv \frac{\sig_{1{\rm m}}^2}{\sig_0\sig_{2{\rm G}}} = \frac{\avg{-\del\,\nabla^2\del_{\rm G}}}{\sqrt{\avg{\del^2}\avg{(\nabla^2\del_{\rm G})^2}}} = \avg{x\mu}\,,
\label{gam}
\end{align}
where we defined the standardised variables 
\be
\mu \equiv \del/\sig_0\qquad{\rm and}\qquad x\equiv-\nabla^2\del_{\rm G}/\sig_{2{\rm G}}\,.
\label{mu-x}
\ee
Note that this definition of \gam\ is similar but not identical to the corresponding one in \citet{bbks86}, since our peak heights are defined using TopHat smoothing.

\subsection{Matching filter scales}
\label{RgRth}
The technical problem which we address in this subsection is how to ensure that the peaks in the Gaussian filtered density field $\del_{\rm G}$ will have an overdensity $\del_{\rm TH}=\delc$ when smoothed with a TopHat, since essentially all measurements in simulations use TopHats only.  In practice, we need a mapping between the scales $R_{\rm TH}$ and $R_{\rm G}$ of the two filters.  All previous work accomplishes this either by matching the volumes:
  $(4\pi/3)\,R_{\rm TH}^3 = (2\pi)^{3/2}\,R_{\rm G}^3$,
or by matching the variances:  
  $\avg{\delta_{\rm TH}^2} = \avg{\delta^2_{\rm G}}$.  
In this second case, the relation between $R_{\rm TH}$ and $R_{\rm G}$ depends on the shape of the power spectrum.  But neither of these conditions guarantee that a peak identified on scale $R_{\rm G}$ satisfies $\del_{\rm TH}=\delc$.  

For this reason, we construct the mapping between $R_{\rm TH}$ and $R_{\rm G}$ by finding that $R_{\rm G}$ (at a given $R_{\rm TH}$) for which $\avg{\del_{\rm G}|\del_{\rm TH}} = \del_{\rm TH}$.  This is equivalent to $\avg{\del_{\rm G}\del_{\rm TH}} = \avg{\del_{\rm TH}^2}$. This definition circumvents the technical complication arising from the variance of the Laplacian of $\del_{\rm TH}$ (which involves a divergent integral).  For the $\Lambda$CDM $P(k)$ we consider in this paper, $R_{\rm G}\approx0.46R_{\rm TH}$ with a mild mass-dependence which we account for, in the range we explore, $3\times10^{10}< m/(h^{-1}\Msun) < 3\times10^{15}$.

Our choice of matching between $R_{\rm TH}$ and $R_{\rm G}$, namely $\avg{\del_{\rm G}|\del}=\del$, means that 
$p(\del_{\rm G}|\del)$ is a Gaussian with mean \del\ and variance $\avg{\del_{\rm G}^2}-s$.  
However, $\avg{\del_{\rm G}^2}\approx s$ to better than $5\%$ and, thus, $p(\del_{\rm G}|\del)\approx\dir(\del_{\rm G}-\del)$ (since the cross-correlation between $\delta_{\rm G}$ and $\delta$ is very close to unity).  This significantly simplifies our ESP analysis for the following reason.  
In principle, the excursion set peaks framework also requires us to keep track of the variable $\xtil\equiv\del'/\sqrt{\avg{\del'^2}}$, where $\del'=\der\del/\der s$. This scalar (isotropic) variable correlates with $\mu$ and $x$, but not with the other (anisotropic) variables used by \citet{bbks86}. Therefore, one should in principle work with the three-dimensional Gaussian vector $(\mu,x,\xtil)$. However, because $p(\del_{\rm G}|\del)\approx\dir(\del_{\rm G}-\del)$, it turns out that setting $\xtil=x$ yields an excellent approximation (notice that this relation is exact for the Gaussian filter).
We have indeed checked that our final answers for $dn/d\ln m$ only change by a few percent 
if we switch between the approximate and exact treatments. In the following, we will therefore always set $\xtil=x$ and only work with the two-dimensional Gaussian vector $(\mu,x)$. 

Even though we smooth the density field with a tophat filter, there is no compelling reason to use such a filter except to make a connection with the spherical collapse approximation. Numerical simulations indeed suggest that, while the Lagrangian volume occupied by the proto-halos is rather compact, it is more diffuse than a tophat window \citep{dwbs08,pdh02,dts13}. However, since a deep understanding of halo collapse is still lacking, we will stick to the tophat (and Gaussian) filters for simplicity.

\subsection{Excursion set peaks with a constant barrier}
\label{constant}

Apart from the fact that the peak height is defined using TopHat filtering, there is no formal change in the derivation by \citet{ps12} of the number density of excursion set peaks for a constant barrier \cite[and this derivation is formally the same as that in][]{aj90}.  In this case we get
\begin{align}
f_{\rm ESP}(\nu) &= V \Cal{N}_{\rm ESP}(\nu)\notag\\
&=(V/V_{\ast})({\rm e}^{-\nu^2/2}/\sqrt{2\pi})\notag\\ 
&\ph{1/V_\ast}\times
 \int_0^\infty\der x\,\frac{x}{\gamma\nu}\,F(x)p_{\rm G}(x-\gam\nu;1-\gam^2)\,,
\label{fESP-const}
\end{align}
where $V=m/\bar\rho=4\pi R^3/3$ is the Lagrangian volume associated with the TopHat smoothing filter, $F(x)$ is the peak curvature function from equation~(A15) of \citet{bbks86} and $p_{\rm G}(y-\bar y;\Sigma^2)$ is a Gaussian distribution in $y$ with mean $\bar y$ and variance $\Sigma^2$.  \cite[As noted in ][$F(x)\neq 1$ reflects the fact that the ESP calculation averages over a special subset of positions.]{ps12}  Since we have been careful to define $\nu$ and hence the mass using TopHat filtering, the mass function follows in the usual way from \eqn{excsetansatz}.  

\begin{figure}
 \centering
 \includegraphics[width=\hsize]{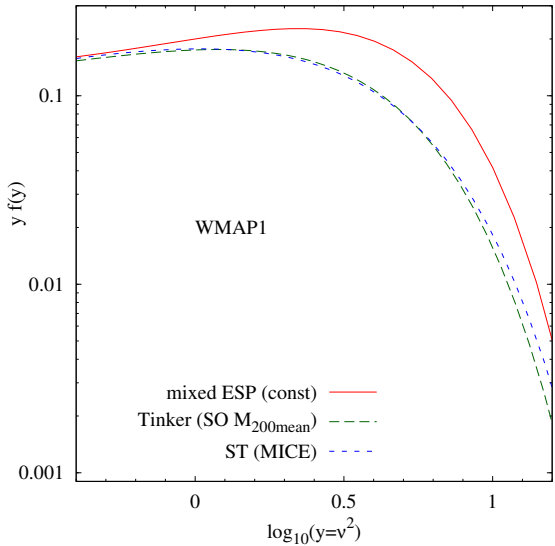}
 \caption{Excursion set peaks mass function for the mixed filtering approach discussed in the text, using a constant barrier (solid red). This is compared with the fit to $N$-body simulations given by \citet{t+08}, appropriate for spherical overdensity halos that enclose $200$ times the mean density of the universe (dashed green). For comparison, we also show the \citet{st99} mass function (short-dashed blue) with their $(q,p)=(0.7,0.26)$, which gives a good fit to the MICE simulations \citep{mice}. The ESP+constant barrier mass function overpredicts the abundance at all scales of interest. All curves used the WMAP1 cosmology -- we have checked that changing the cosmology does not significantly alter the comparison.
   }
 \label{vfv-const}
\end{figure}

\fig{vfv-const} compares the ESP result for a constant barrier $B(s)=\delc=1.686$ (solid red) with the fit to $N$-body simulations from \cite{t+08} (dashed green). For the latter we used parameters from Table 2 of \cite{t+08} appropriate for halos identified using the spherical overdensity (SO) definition at $200$ times the mean density of the universe. We used cosmological parameters for the WMAP1 cosmology mentioned in the Introduction, for which the \cite{t+08} function gives a good fit (c.f. their Table 2). Later we will also show results for the WMAP3 cosmology which was also explored by \citet{t+08} -- changing cosmology does not affect our results significantly. For comparison we also show the functional form of \cite{st99} (short-dashed blue) with their $(q,p)=(0.7,0.26)$ which gives a good fit to the MICE simulations \citep{mice}.

Clearly, equation~(\ref{fESP-const}) predicts far too many halos at all interesting scales, including the high mass end ($\nu^2=10$ corresponds to $m=1.3\times10^{15}h^{-1}\Msun$ for this cosmology).  

\subsection{Beyond the spherical-collapse constant barrier}
It is tempting to attribute the mismatch shown in Figure~\ref{vfv-const} to the effects of non-spherical collapse.  While these effects are certainly important at smaller mass scales, it is not obvious that they also matter at the largest scales, where the spherical collapse approximation should be quite accurate. To see that they are in fact relevant, note that the triaxial collapse model predicts that the critical density for collapse scales approximately as 
\be
 \delta_{\rm ec} \approx \delc[1 + 0.4 (\sig_0/\delc)^{1.2}]
 \label{dec}
\ee 
\citep{smt01}.  Simulations indeed show that this expression for $\delta_{\rm ec}$ provides a reasonable approximation to the mass dependence of the critical density required for collapse \citep{smt01}.  However, this implies that when $\nu = \delc/\sig_0$ is as large as $\sim 4$, then departures from the spherical \delc\ are already of the order of ten percent.  Therefore, we must include deviations from the spherical collapse barrier in order to have a fair comparison with the \citet{t+08} simulations. 

Before explaining how this can be done, we note that, although equation~(\ref{dec}) is a reasonable approximation to the mass dependence of the critical density required for collapse \citep{smt01}, there is substantial scatter around this value \citep{dwbs08, rks09}.  This scatter has been quantified by \citet{rktz09}\footnote{We use the results of \citet{rktz09} rather than those of \citet{dwbs08} because the latter only show a plot whereas Robertson et al. provide a description of the full probability distribution. Both groups find a similar mean and r.m.s. for the barrier heights, however, which increase approximately linearly with $\sig_0$.} who found that, over the range $\log_{10}(\nu^2)\in(-0.3,0.9)$, the value of $B$ was Lognormally distributed around a mean value of $\simeq 1.51 + 0.49\sig_0$ with variance $(0.09\sig_0^2)$ (c.f. their Table 1 and figure 3).  The physical origin of this scatter is unclear.  Triaxial collapse predicts some scatter \citep{smt01, dts13}, which is related to the statistics of the tidal field, but the measured scatter is considerably larger. In any case, the physical origin of this scatter is unimportant for what follows - though we believe that this is an issue which deserves further investigation.

\subsection{Excursion set peaks with a square-root barrier and scatter}
\label{sqrtscatt}
To include the mean trend and scatter in barrier heights shown by \citet{smt01} and quantified by \citet{rktz09}, we will adopt the following model, first discussed in \citet{pls12}.  We consider a family of deterministic ``square-root'' barriers $B$ given by 
\be
 B = \delc + \beta\sqrt{s}\,,
\label{sqrtbarrier}
\ee
with $\delc=1.686$ and $\beta$ a stochastic variable.  Therefore, if $f_{\rm ESP}(\nu|\beta)$ denotes the ESP solution for a given value of $\beta$, then our full solution will be 
\be
f_{\rm ESP}(\nu) = \int \der\beta\,p(\beta)\,f_{\rm ESP}(\nu|\beta).
\label{fESP}
\ee
Below we use simulations to motivate our choice of $p(\beta)$, and provide an explicit expression for $f_{\rm ESP}(\nu|\beta)$.  Still, it is easy to understand qualitatively what happens.  

\begin{figure}
 \centering
 \includegraphics[width=\hsize]{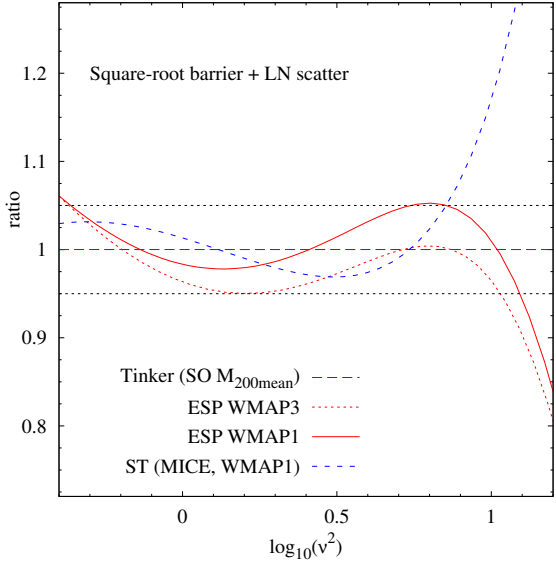}
 \caption{Ratio of the ESP mass function for a square-root barrier with scatter in the barrier slope (equation~\ref{fESP}) to the \citet{t+08} fit, for WMAP1 (solid red) and WMAP3 (dotted red) cosmologies. In each case we set $\delc=1.686$ and $p(\beta)$ to be Lognormal with $\avg{\beta}=0.5$ and ${\rm Var}(\beta)=0.25$, for the reasons discussed in the text. For comparison we also show the ratio of the MICE fit with the Tinker \etal\ fit for the WMAP1 cosmology (short-dashed blue). Horizontal dotted lines mark $5\%$ deviations from the Tinker \etal\ fit; the typical scatter in their measurements in this mass range is $5$-$10\%$.
   }
 \label{ratio}
\end{figure}

For excursion set peaks, the square root barrier \eqref{sqrtbarrier} with $\beta>0$ dramatically lowers the halo counts at large masses \cite[Figure~3 in][]{ps12}. While this goes in the right direction, introducing scatter in the value of $\beta$ will counteract this decrease. This is because a scatter in $\beta$ induces a scatter in the predicted mass of the halo. The steepness of the mass function implies that the scatter preferentially moves small mass objects to larger masses (the same effect that leads to Eddington bias), which in turn increases the abundance at large masses.

The importance of the scatter depends on the distribution of $\beta$, our choice for which is motivated by the Lognormal distribution of $B$ seen in simulations for halos of a fixed mass. Notice that what we need as input is the `prior' distribution $p(\beta)$, knowing which our analysis then predicts $f_{\rm ESP}(\nu)$ using \eqn{fESP}. However, the distribution reported by \citet{rktz09} corresponds to the \emph{conditional} distribution $p(\beta|\nu)$ which we can only construct \emph{a posteriori} using the relation 
\be
 p(\beta|\nu) = f_{\rm ESP}(\nu|\beta)p(\beta)/f_{\rm ESP}(\nu)\,,
\label{pbetanu}
\ee
which is a simple application of Bayes' theorem. 

In practice, we choose $p(\beta)$ to be Lognormal with mean and variance chosen such that the mass function $f_{\rm ESP}(\nu)$ as well as the conditional distribution $p(\beta|\nu)$ are in reasonable agreement with the results of simulations. Having done this, we will see later that the clustering of halos is also correctly predicted by our model, which is a nontrivial test of the formalism. We set $\avg{\beta}=0.5$ and ${\rm Var}(\beta)=0.25$ for the Lognormal $p(\beta)$; we have checked that varying these numbers within a $\sim10\%$ range does not affect our results significantly. We avoided fitting these numbers more accurately because, as mentioned earlier, their physical origin is at present unclear and a more accurate fit would not lead to any physical insights.  Moreover, recent work, in which halos were identified using an ellipsoidal rather than spherical halo finder, suggests that the mean value is closer to $1.68\,(1 + 0.2\sig_0)$ with variance around this mean of $0.04\sigma_0^2$ \citep{dts13}.  

\begin{figure}
 \centering
 \includegraphics[width=\hsize]{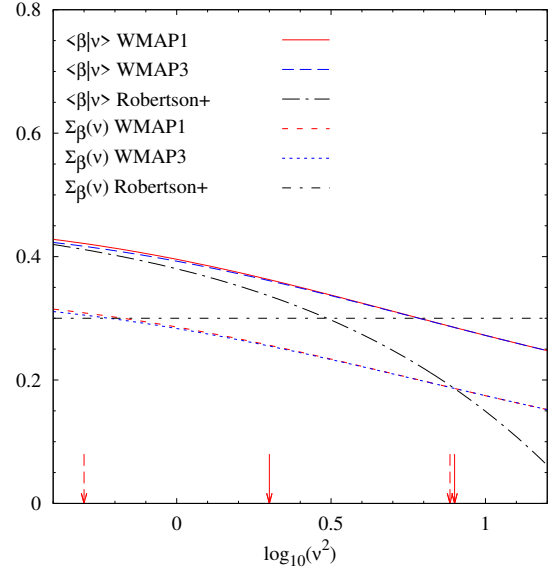}
 \caption{The conditional mean $\avg{\beta|\nu}$ and scatter $\Sigma_\beta(\nu)\equiv\sqrt{{\rm Var}(\beta|\nu)}$ of the distribution $p(\beta|\nu)$ (equation~\ref{pbetanu}) for the WMAP1 (red) and WMAP3 (blue) cosmologies. The black curves show the corresponding quantities measured by \citet{rktz09}. The ESP curves used $\delc=1.686$, $f_{\rm ESP}(\nu|\beta)$ from \eqn{fESP-beta} and assumed $p(\beta)$ to be Lognormal with $\avg{\beta}=0.5$ and ${\rm Var}(\beta)=0.25$. The curve for $\avg{\beta|\nu}$ from \citet{rktz09} assumed the `linear' fit from their Table 1, $B_{\rm fit} = 1.506 + 0.487\sig_0$, and shows $\avg{\beta|\nu}=(B_{\rm fit}-1.686)/\sig_0$. The arrows indicate the approximate range over which Robertson et al. performed these fits for $\avg{\beta|\nu}$ (dashed) and $\Sigma_\beta(\nu)$ (solid). The ESP predictions are within $30\%$ of the measured quantities.
   }
 \label{betanu}
\end{figure}

For square-root barriers with fixed $\beta$ \citep[equation 19 of][for the barrier in equation~\ref{sqrtbarrier} above]{ps12},
\begin{align}
f_{\rm ESP}(\nu|\beta) &= (V/V_\ast) ({\rm e}^{-(\nu+\beta)^2/2}/\sqrt{2\pi})\notag\\
&\ph{V/V_\ast}\times
 \int_{\beta\gam}^\infty \der x\,\frac{x-\beta\gam}{\gamma\nu} F(x)\notag\\
&\ph{V/V_\ast\times\int}\times p_{\rm G}(x-\beta\gam - \gam\nu;1-\gam^2)\,.
\label{fESP-beta}
\end{align}
We have assumed $\beta>0$, which is true for the Lognormal model we will consider. If $\beta$ is allowed to take negative values as well, then the lower limit of the integral over $x$ must be replaced by $x_{\rm min} = {\rm max}\{0,\beta\gam\}$, so as to enforce the peak constraint. 
While there is no closed form expression for $f_{\rm ESP}(\nu)$, the integrals involved can be easily computed numerically. 

\fig{ratio} shows the results for the WMAP1 (solid red) and WMAP3 (dotted red) cosmologies. Note that the value of $\sigma_8$ is $\sim10\%$ smaller in the WMAP3 cosmology. We plot the ratio of the ESP prediction in \eqn{fESP} and the \citet{t+08} fit. For comparison we also indicate the ratio of the MICE fit with the Tinker \etal\ fit (for the WMAP1 cosmology) as the short-dashed blue curve. The horizontal dotted lines mark $5\%$ deviations from the Tinker \etal\ fit; the typical scatter in their measurements in this mass range is $5$-$10\%$ \citep[c.f. figure 6a of][]{t+08}.  

\fig{betanu} shows the predicted mean $\avg{\beta|\nu}$ and scatter $\Sigma_\beta(\nu)$ of the conditional distribution $p(\beta|\nu)$ (equation~\ref{pbetanu}) for the WMAP1 (red) and WMAP3 (blue) cosmologies. The black curves show the corresponding quantities measured by \citet{rktz09}, assuming the fit $B_{\rm fit} = 1.506 + 0.487\sig$ from their Table 1 and using $\avg{\beta|\nu}_{\rm Robertson+}=(B_{\rm fit}-1.686)/\sig$ in order to compare with our square-root barrier model in \eqn{sqrtbarrier}. The arrows indicate the approximate range over which Robertson et al. performed these fits for $\avg{\beta|\nu}$ (dashed) and $\Sigma_\beta(\nu)$ (solid). The ESP predictions are within $30\%$ of the measured quantities. Together with \fig{ratio}, this shows that our model works rather well. 

\section{Implications for averages over all, rather than a special subset of positions} \label{allwalks}

The ESP model of the previous section is the first to self-consistently integrate the complex physics of non-spherical collapse with the fact that the physics is almost certainly simplest around the center-of-mass of the proto-halo.  Although the agreement with the halo counts represents a significant and nontrivial success, and the next section shows that the predicted spatial distribution is also accurate, it makes a number of other testable predictions which are related to the logic of the excursion set approach.  

For example, one can construct the center-of-mass random walk associated with each halo and explicitly check, object by object, whether or not the corresponding barrier was crossed at a larger scale. This is, in fact, precisely the kind of test which was performed by \cite{smt01}.  Revisiting this test is left to future work.

\subsection{Distribution of $\delta$ for all particles in proto-halos of a given mass}
We remarked in the Introduction that our ESP model explicitly averages only over special positions in the universe (see text following equation~\ref{fESP-const} for exactly where this happens).  If we treat these positions as the fundamental quantities -- for which the physics of collapse is most easily applied -- then by accounting for their initial density profiles, it is straightforward to estimate what would have happened had we averaged over all positions.  Because our model has been calibrated for center of mass positions solely, predicting the right distribution of $\delta$ values within each proto-halo provides another test of our approach.  We illustrate the idea below.

We begin by noting that because $\delta_{\rm pk} \equiv \delta_c + \beta\sigma_0$, 
the distribution of $\beta$ at fixed $\nu\equiv\delta_c/\sigma_0$ 
implies a distribution of $\delta_{\rm pk}$:  
\begin{equation}
 p(\delta_{\rm pk}|S)\,{\rm d}\delta_{\rm pk} = p(\beta|\nu)\,{\rm d}\beta .
\end{equation}
Recall that fixed $\nu$ means fixed halo mass $m$, smoothing scale $R$ 
or variance $S$, and simulations suggest $p(\beta|\nu)$ is approximately 
Lognormal.  

\begin{figure}
 \centering
 \includegraphics[width=\hsize]{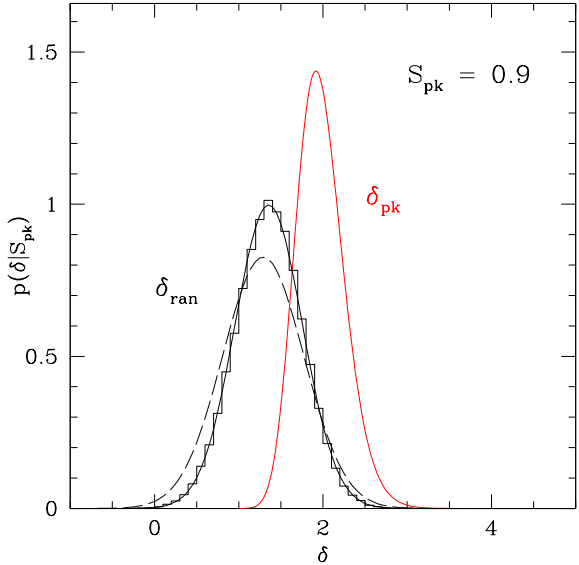}
 \caption{Predicted distribution of $\delta$ for all particles in 
          proto-halos of a given mass.  Histogram shows our model, 
          equation~(\ref{pd1pk}), which uses as input the smooth 
          Lognormal distribution shown for $p(\delta_{\rm pk}|S_{\rm pk})$.  
          Our model is well approximated by a Gaussian (solid curve); 
          it has a slightly different mean and variance than do the 
          measurements in simulations (the dashed Gaussian).}
 \label{pdeltaPk}
\end{figure}

Now, suppose that there is an excursion set peak of height $\delta_{\rm pk}$ identified on smoothing scale $R_{\rm pk}$.  Let $p(\delta|r,\delta_{\rm pk},S_{\rm pk})$ denote the probability that the overdensity on smoothing scale $R_{\rm pk}$ at distance $r$ from this peak is $\delta$.  This distribution is Gaussian, with mean and variance given by equations~(7.8--7.10) in \cite{bbks86}.  Therefore, if we choose a random position within the set of all initial peak-patches of scale $R_{\rm pk}$, then it will have overdensity $\delta$ with probability 
\begin{equation}
 p(\delta|S_{\rm pk}) = \int d\delta_{\rm pk}\,p(\delta_{\rm pk}|S_{\rm pk}) \,
      \int_0^R 3\frac{{\rm d}r}{r}\, \frac{r^3}{R^3}\,p(\delta|r,\delta_{\rm pk},S_{\rm pk}) .
 \label{pd1pk}
\end{equation}
This expression is particularly easy to evaluate using Monte-Carlo methods:  generate $(r/R)^3$ and $\delta_{\rm pk}$ from uniform and Lognormal distributions respectively, and then generate a Gaussian variate whose mean and variance depend on $r/R$ and $\delta_{\rm pk}/\sqrt{S_{\rm pk}}$, following \cite{bbks86}.  The histogram in Figure~\ref{pdeltaPk} shows the result when $S_{\rm pk}=0.9$ (this corresponds to halo masses of $7.8\times 10^{13}h^{-1}M_\odot$).  The smooth solid curve which provides an excellent fit shows a Gaussian with the same mean and variance as $\delta$.  We have included it because the mean and variance of $p(\delta|r,\delta_{\rm pk},S_{\rm pk})$ are known analytically \citep[from][]{bbks86}, so the mean and variance of $p(\delta|S_{\rm pk})$ are simply related to volume averages of correlation functions and the mean and variance of $p(\delta_{\rm pk}|S_{\rm pk})$.  

The dashed curve shows a Gaussian distribution with mean $(1.67 + 0.075\,S_{\rm pk})/1.35$ and variance $(0.35/1.35)\,S_{\rm pk}$, which describes the measurements in simulations of this quantity rather well \citep{arsc13}.  In effect, our model is trying to predict this dashed curve.  Because we use as input the Lognormal distribution of peak heights $p(\delta_{\rm pk}|\nu)$ that is seen in simulations (shown as the other solid curve which peaks at larger $\delta$:  it has mean $\avg{\delta_{\rm pk}|S_{\rm pk}} = 1.506 + 0.487\sqrt{S_{\rm pk}}$, and variance $0.09\,S_{\rm pk}$), our model has \emph{no} free parameters, so it is perhaps remarkable that the model does as well as it does.  Although our prediction is narrower than the measured distribution (predicted variance is $0.18\,S_{\rm pk}$ whereas the measured value is $0.26\,S_{\rm pk}$) its mean, 1.35, is within five percent of the measured value of 1.29.  

\begin{figure}
 \centering
 \includegraphics[width=\hsize]{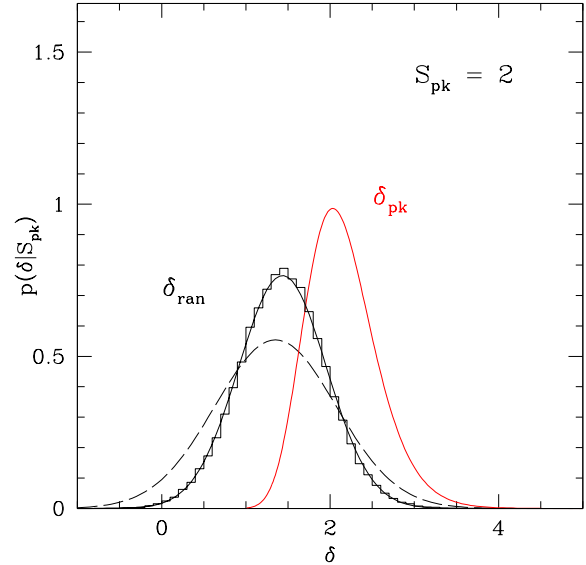}
 \caption{Same as previous Figure, but now for $S_{\rm pk}=2$.  }
 \label{pdeltaPk2}
\end{figure}

Figure~\ref{pdeltaPk2} shows a similar analysis for $S_{\rm pk}=2$, which corresponds to lower mass halos ($7.8\times 10^{12}h^{-1}M_\odot$) for which the assumption that halos form from peaks is less secure; clearly, the discrepancies between the model and the measurements are qualitatively similar even at these small masses.  The predicted mean and variance are $1.43$ and $0.136\,S_{\rm pk}$, whereas the measured values are 1.36 and $0.26\,S_{\rm pk}$.

%S0 = 0.89;  nu = 2.08
%<d1pk> = 1.36
%var(d1pk)/S = 0.11;  rms/sqrtS = 0.33
%<d1DEUS> = 1.29
%<d1mc> = 1.35; rmsMC = 0.42 sig; varMC = 0.18 S

%S0 = 2.11;  nu = 1.52
%<d1pk> = 1.44
%var(d1pk)/S = 0.12;  rms/sqrtS = 0.35
%<d1DEUS> = 1.36
%d1mc> = 1.43; rmsMC = 0.37 sig; varMC = 0.136 S

\subsection{Distribution of predicted mass for all particles in proto-halos of a given mass}
The lower mean value implies that the mass which the excursion set approach predicts for the particles which are not at the center of mass should be systematically smaller than the mass of the halo in which they end up -- in qualitative agreement with one of the findings of \citet[][see their Figures~2-4]{smt01}.  

To see that this is indeed the case, we can use our ESP approach (walks centered on peaks in the initial field) to model what the usual excursion set estimate (walks centered on all positions in the initial field) really represents.  Recall that, for the statistics of all positions, we know that if $\delc$ is independent of $m$ then the excursion set prediction for the mass fraction in objects with mass $m(s)$ is $sf(s) = \nu\,\exp(-\nu^2/2)/\sqrt{2\pi}$ with $\nu^2 = \delta_c^2/s$.  (Strictly speaking, to compare most directly with \cite{smt01} we should account for the distribution in $\delc$ values at each $s$; but the constant deterministic \delc\ discussion below captures the gist of the argument.  Also, our expressions below are appropriate for a sharp-$k$ filter, but the logic, and so the qualitative trends, do not depend on this choice.)  

In our ESP model, this fraction is given by taking each peak of mass $M$, then accounting for the distribution of excursion set predictions for $s(m)>S(M)$ which come from all the other mass elements which make up $M$ (rather than just the one mass element at the center of mass), and integrating the distribution of $S$ over the range $0<S<s$.  I.e., 
\be
 sf(s) = \int_0^s \frac{{\rm d}S}{S}\,Sf_{\rm ESP}(S)\, sf(s|S).
 \label{ESPtoPS}
\ee
where
\be
 Sf_{\rm ESP}(S) = \frac{M}{\bar\rho}\,\frac{{\rm d}n_{\rm ESP}(M)}{{\rm d}\ln M}\,
                 \left|\frac{{\rm d}\ln M}{{\rm d}\ln S}\right| 
 \label{ESP}
\ee
(from equation~\ref{excsetansatz}) and $f(s|S)$ is the excursion set prediction for the mass fraction of $M$ which is incorrectly predicted to instead have mass $m = M(s)$ instead of $M(S)$.  Because we know $sf(s)$ as well as $n_{\rm ESP}(S)$ we can solve (by numerical back substitution) for $f(s|S)$.  However, we can get a simple, intuitive estimate of the result as follows.  

For constant \delc, Figures~1 and~2 in \citet{ps12} shows that $Sf_{\rm ESP}(S)$ is rather well-approximated simply by $\nu f(\nu) = \nu\,\exp(-\nu^2/2)/\sqrt{2\pi}$ with $\nu^2 = (\Xi\delta_c)^2/S$ for some $\Xi<1$.  (Strictly speaking, they used the Sheth-Tormen functional form for $\nu f(\nu)$; however, at high masses, this form reduces to the simpler one given here with $\Xi=\sqrt{0.7}$.)  Since this is the same functional form for $S$ that appears on the left hand side of equation~\eqref{ESPtoPS} for $s$, it is easy to check that $sf(s|S)$ should also be the same function of $\nu$, but with $\nu^2 = (\delta_c -  \Xi\delta_c)^2/(s-S)$.  

Thus, this calculation shows explicitly that the average over all walks will yield $f$ which has a similar functional form to the one for the specially chosen subset, but the two will have different values of $\delta_c$.  Moreover, the distribution of predicted masses will be very different from a delta function centered on $M$.  Rather, it will look just like the conditional excursion set prediction for the fraction of all walks which first cross \delc\ on scale $s$ given that they are known to pass through $\Xi\delc < \delc$ on scale $S$ before first crossing \delc.  This latter point quantifies the explanation given by \citet{smt01} when addressing the concerns raised by \citet{w96}.  It also expands on a point first made by \citet{ps12}: the ESP framework naturally accounts for the fact that one must rescale $\nu\to\sqrt{q}\nu$ with $q\sim0.7$ if one wants to use the usual excursion set expressions (based on the statistics of all positions) to model halo counts -- but this rescaling is not necessary if one uses the statistics of the special positions for which the physics is simplest.  Of course, because the physics of collapse refers to collapse around the halo centers of masses, and not around random positions in space, this discussion highlights the dangers of interpreting the barrier height in the random walk calculation with that for the physics of collapse around special positions in space.

\section{Halo bias}
\label{bias}
In this section we derive expressions for the predicted clustering of halos in our model.

In the excursion set approach, the clustering of halos relative to the initial dark matter field -- or `Lagrangian halo bias' -- is captured by the conditional distribution $f(s|\delo,\So)$ of walks that first cross the barrier at some small scale $s$ having previously passed through the value \delo\ at some large scale \So. As discussed by \citet{mps12} and \citet{ps12}, a convenient way of organising this bias, for both excursion sets and excursion set peaks, is through cross-correlations of the halo density $\avg{\rhoh|\delo} = f(\nu|\delo,\So)/f(\nu)$ with Hermite polynomials in the matter overdensity \delo\:
\begin{align}
b_n &\equiv \So^{-n/2}\avg{\rho_hH_n(\delo/\sqrt{\So})}\notag\\
&= \So^{-n/2}\int_{-\infty}^{\infty}\der\delo p_{\rm G}(\delo;\So)
\avg{\rhoh|\delo} H_n(\delo/\sqrt{\So})\,, 
\label{bn-def}
\end{align}
where $H_n(x)={\rm e}^{x^2/2}(-d/dx)^n {\rm e}^{-x^2/2}$ are the ``probabilist's'' Hermite polynomials, and the bias parameters $b_n$ can be shown to have the structure
\be
b_n =\left(\frac{\Sc}{\So}\right)^n \sum_{r=0}^n \binom{n}{r}
b_{nr}\epc^r\,,  
\label{bn-expand}
\ee
where
\begin{align}
\Sc &= \int\der\ln k\,\Del^2(k)W(kR)W(kR_0)\,,\notag\\
\epc &= 2\,\der\ln\Sc/\der\ln s\,
\label{Scepc}
\end{align}
are cross-correlations between the small and the large scale. 

The calculation of bias therefore requires the conditional mass function of excursion set peaks. Introducing scatter in the barrier parameter $\beta$ is straightforward, as we discuss below.
The ESP conditional multiplicity function for a square-root barrier at fixed $\beta$ \emph{and} a fixed density \delo\ at scale \So\ is given by
\begin{align}
f_{\rm ESP}(\nu|\beta,\delo) &= (V/V_\ast) p(B/\sqrt{s}|\delo) \notag\\
&\ph{V_\ast}\times
\frac1{\gam\nu} \int_{\beta\gam}^\infty \der x\,(x-\beta\gam)F(x)\notag\\
&\ph{V_\ast p(B/\sqrt{s}|\delo)}\times
p(x|B/\sqrt{s},\delo)\,,
\label{fESPcond-sqrt-full}
\end{align}
where $p(\mu|\delo)$ and $p(x|\mu,\delo)$ are 1-dimensional conditional Gaussian distributions that can be written using the covariance matrix of the trivariate Gaussian in $(\mu,x,\delo)$. 

The bias coefficients $\delc^nb_n$ can be shown to be algebraically equivalent to the Taylor coefficients of the expansion of \eqn{fESPcond-sqrt-full} in powers of $\bar\del_0\equiv\delo/\delc$, in a specific limit \citep{mps12,ps12}. This leads to a straightforward prescription for calculating these quantities. In particular, one must evaluate \eqn{fESPcond-sqrt-full} after formally setting $\tilde\cb=0$, where $\tilde\cb$ is the matrix that one must subtract from the unconditional covariance matrix of the variables $(\mu,x)$ to obtain the covariance matrix of the conditional Gaussian $p(\mu,x|\delo)$. The resulting conditional multiplicity is
\begin{align}
&f_{\rm ESP}(\nu|\beta,\delo,\tilde\cb=0) \notag\\
&\ph{1}= (V/V_\ast) ({\rm e}^{-(\nu+\beta-\bar\del_0(\Sc/\So)\nu)^2/2}/\sqrt{2\pi})\notag\\
&\ph{1V_\ast} \times%
\frac1{\gam\nu} \int_{\beta\gam}^\infty \der x\,(x-\beta\gam)F(x)\notag\\
&\ph{1V_\ast\int\der x x F(x)} \times
p_{\rm G}(x-\beta\gam - \gam\nu_1;1-\gam^2)\,,
\label{fESPcond-sqrt-ctileq0}
\end{align}
where we have defined $\nu_1\equiv \nu(1- \bar\del_0(\Sc/\So)(1-\epc))$, and \Sc\ and \epc\ were defined in \eqn{Scepc}.

Taylor expanding \eqn{fESPcond-sqrt-ctileq0} in powers of $\bar \del_0$, the bias parameters at fixed $\beta$ can be written as
\begin{align}
\frac{\delc b_1(\nu,\beta)}{(\Sc/\So)} &= \mu_1(\nu,\beta) + (1-\epc)\lam_1(\nu,\beta)\,,
\label{dcb1-beta}\\
\frac{\delc^2 b_2(\nu,\beta)}{(\Sc/\So)^2} &= \mu_2(\nu,\beta) + 2(1-\epc)\mu_1(\nu,\beta)\lam_1(\nu,\beta)\notag\\ 
&\ph{\mu_2}
+ (1-\epc)^2\lam_2(\nu,\beta)\,,
\label{dc2b2-beta}
\end{align}
where, using $\Gam\equiv\gam/\sqrt{1-\gam^2}$,
\begin{align}
\mu_n(\nu,\beta) &= \nu^nH_n(\nu+\beta) \,,
\label{mu_n-beta}\\
\lam_n(\nu,\beta) &= (-\Gam\nu)^n\avg{H_n\left.\left(y-\beta\Gam-\Gam\nu\right)\right|\nu,\beta}_y\,,
\label{lam_n-beta}
\end{align}
and for some function $h(y,\nu,\beta)$, we have
\begin{align}
&\avg{h(y,\nu,\beta)|\nu,\beta}_y \notag\\
&= \frac{\int_{\beta\Gam}^\infty\der y\,(y-\beta\Gam)F(y\gam/\Gam)p_{\rm G}(y-\beta\Gam-\Gam\nu;1)h(y,\nu,\beta)}{\int_{\beta\Gam}^\infty\der y\,(y-\beta\Gam)F(y\gam/\Gam)p_{\rm G}(y-\beta\Gam-\Gam\nu;1)}.
\label{avgx-beta}
\end{align}
Marginalising over $\beta$ gives the bias parameters as
\begin{align}
\delc b_1(\nu) &= \left(\frac{\Sc}{\So}\right)\bigg[\avg{\mu_1|\nu} + (1-\epc)\avg{\lam_1|\nu} \bigg]\,,
\label{dcb1}\\
\delc^2 b_2(\nu) &= \left(\frac{\Sc}{\So}\right)^2\bigg[\avg{\mu_2|\nu} + 2(1-\epc)\avg{\mu_1\lam_1|\nu}\notag\\
&\ph{\left(\frac{\Sc}{\So}\right)^2\bigg[\mu_2\bigg]}
 + (1-\epc)^2\avg{\lam_2|\nu}\bigg]\,,
\label{dc2b2}
\end{align}
where, for some function $g(\nu,\beta)$, we have
\be
\avg{g|\nu} = \int\der\beta\,p(\beta|\nu)g(\nu,\beta)\,,
\label{avgbeta}
\ee
where $p(\beta|\nu)$ was given in \eqn{pbetanu}.
\begin{figure}
 \centering
 \includegraphics[width=\hsize]{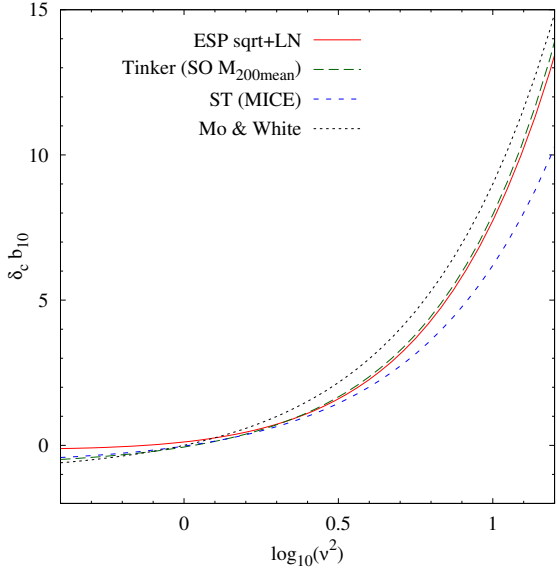}
 \caption{Scale-independent (peak-background split) linear Lagrangian halo bias $\delc b_{10}$ in the ESP framework for a square-root barrier with Lognormal scatter (solid red). This curve is in excellent agreement with the fit to $N$-body simulations from \citet{t+10}, the simulations being the same as those for which \fig{ratio} showed a mass function comparison (see text for a discussion). The dotted black curve is the spherical collapse prediction $\nu^2-1$ \citep{mw96}, and the short-dashed blue curve is the peak-background split prediction associated with the mass function fit to the MICE simulations.
   }
 \label{dcb10}
\end{figure}

The presence of \epc\ in these expressions for real-space bias is an indication of $k$-dependent Fourier-space bias, as discussed at length by \citet{mps12}\footnote{The factor $(\Sc/\So)^n$ is a consequence of the fact that the excursion set calculation lives in real space rather than Fourier space, and would be present even if the Fourier-space bias were constant.}. At large scales $R_0\to \infty$, $\epc\to0$. Consequently, the coefficients $b_{n0}$ in \eqn{bn-expand} correspond to the large scale ($k\to0$) Fourier-space bias that is routinely measured in simulations. Further, \citet{mps12} pointed out that $b_{n0}$ correspond to the usual peak-background split expressions for bias, $b_{n0} = f(\nu)^{-1}(-\p/\p\delc)^nf(\nu)$. In other words, the peak-background split bias parameters correspond to taking the large scale limit $R_0\to\infty$, or equivalently, setting $\epc\to0$ and $\Sc/\So\to1$ in the coefficients $b_n$. 

\fig{dcb10} shows our ESP prediction for the linear peak-background split bias $\delc b_{10}$ (i.e., setting $\epc=0$, $\Sc/\So=1$ in equation~\ref{dcb1}) as the solid red curve. The dotted black curve is the spherical collapse prediction $\nu^2-1$ \citep{mw96}. The dashed green curve shows the fit presented by \citet{t+10} for the same simulations analysed by \citet{t+08}. We have checked that in this case the results for the WMAP1 and WMAP3 cosmologies are nearly identical, so we only show the results for WMAP3. The ESP curve matches the $N$-body fit very well at large masses, which is where the ESP framework is expected to work well. Note that the curve for \citet{t+10} is not simply the logarithmic derivative of the mass function calibrated by \citet{t+08}; that derivative would underestimate the large scale bias by up to $10\%$ \citep{mss10,t+10}. (For comparison, the short-dashed blue curve shows the peak-background split prediction for the MICE mass function.)  

\begin{figure}
 \centering
 \includegraphics[width=\hsize]{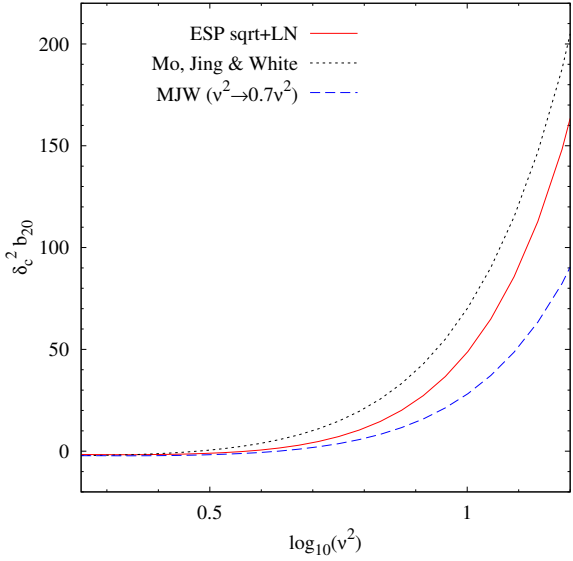}
 \caption{Same as previous Figure, but now for the second order bias coefficient.  In this case, there are no measurements from simulations with which to compare. }
 \label{dc2b20}
\end{figure}

\fig{dc2b20} shows the corresponding result for the quadratic peak-background split coefficient $\delc^2b_{20}$ for ESP (solid red) and the spherical collapse result $\nu^2(\nu^2-3)$ \cite*[dotted black]{mjw97}. The dashed blue curve shows the \citet{mjw97} expression after rescaling $\nu\to\sqrt{0.7}\nu$.  In this case, there are no measurements in simulations with which to compare -- this is the subject of ongoing work -- but we have included the Figure to show that $b_{20}$ behaves qualitatively like $b_{10}$, transitioning from the usual (sharp-$k$ excursion set) expression rescaled by $\nu\to\sqrt{0.7}\nu$ at $\nu\sim 1$ to lying above this rescaled value at larger $\nu$.

These plots illustrate another point made by \citet{ps12}: the ESP framework naturally accounts for the fact that the measured large scale bias deviates from that predicted by a naive application of the peak-background split to a mass function that was fit using the rescaled variable $\nu\to\sqrt{q}\nu$ with $q\sim0.7$.

\subsection{Reconstructing the peak-background split parameters}

In this subsection we revisit the idea of reconstructing the peak-background split parameters $b_{n0}$ from suitable real-space measurements of the parameters $b_n$ \citep{mps12}. In particular, we wish to address some subtleties that arise due to the stochasticity of the barrier height.

The theoretical quantities plotted in Figures~\ref{dcb10} and~\ref{dc2b20} required us to set $\epc\to0$ and $\Sc/\So\to1$ by hand. In Fourier space, this would correspond in practice to taking the limit $k\sim1/R_0\to0$, which is how \citet{t+10} e.g. derived their fit for the linear bias. \citet{mps12}, on the other hand, pointed out that the theoretical predictions for $\delc^n b_n$ at \emph{finite} $R_0$ are equivalent to an average over the Hermite-transformed initial density field $H_n(\delo/\sqrt{\So})$ at the centers of halos (equation~\ref{bn-def}). To be precise, if there are $N$ halos in a given mass bin, one estimates the bias parameters at this mass as
\be
b_n^{\rm (msd)} = \So^{-n/2}\frac1N\sum_{i=1}^{N}H_n(\del_{0i}/\sqrt{\So}), 
\label{bn-msd}
\ee
where $\del_{0i}$ is the Lagrangian density contrast smoothed on scale $R_0$ and centered on the $i^{\rm th}$ halo in the bin, and \So\ is the linearly extrapolated variance at $R_0$.

If we use a deterministic barrier in our theoretical model for the bias, then these measurements can be explicitly converted into estimates of the peak-background split parameters $b_{n0}$ \citep{mps12,ps12}. To see this, note that for the square-root barrier with fixed $\beta$, we can use \eqns{dcb1-beta} and~\eqref{dc2b2-beta} to write the peak-background split parameters in terms of $b_1(\nu,\beta)$ and $b_2(\nu,\beta)$ as
\begin{align}
\delc b_{10}(\nu,\beta) &= \mu_1 + \lam_1\notag\\
&= \frac1{(1-\epc)}\bigg[\frac{\delc b_1(\nu,\beta)}{(\Sc/\So)} - \epc\mu_1 \bigg]\,,
\label{dcb10-beta}\\
\delc^2 b_{20}(\nu,\beta) &= \mu_2 + 2\mu_1\lam_1 + \lam_2\notag\\
&= \frac1{(1-\epc)^2}\bigg[\frac{\delc^2 b_2(\nu,\beta)}{(\Sc/\So)^2}\notag\\ 
&\ph{(1-\epc)^2\bigg[\mu_2\bigg]}
- 2\epc\mu_1\left(\frac{\delc b_1(\nu,\beta)}{(\Sc/\So)} - \mu_1\right)\notag\\
&\ph{(1-\epc)^2\bigg[2\epc\mu_1\lam_1\bigg]}
- \epc(2-\epc)\mu_2 \bigg]\,,
\label{dc2b20-beta}
\end{align}
which reduce to the expressions in equations~(44) and~(47) of \citet{mps12} when $\beta=0$.

If $\beta$ is stochastic (as we have seen it must be), then the measurements on the r.h.s. of \eqn{bn-msd} correspond to the marginalised bias coefficients -- $b_n^{\rm (msd)}\leftrightarrow\avg{b_n(\nu,\beta)|\nu}$ -- since we did not keep track of the value of $\beta$ for each halo. For the linear coefficient $b_{10}$ this does not pose any problem, and we can estimate this coefficient using $\avg{\mu_1|\nu}$, which is calculable, and the measured linear bias, 
\begin{align}
\delc b_{10}^{\rm (msd)}(\nu) &= \frac1{(1-\epc)}\bigg[\frac{\delc b_1^{\rm (msd)}}{(\Sc/\So)} - \epc\avg{\mu_1|\nu} \bigg]\,.
\label{dcb10-avg}
\end{align}
For $b_{20}$, however, we must deal with a term involving $\avg{b_1(\nu,\beta)\mu_1(\nu,\beta)|\nu}$. In practice, measuring this term would involve knowing the value of $\beta$ for each halo. While this is do-able in an $N$-body simulation, it makes the reconstruction technique very cumbersome. In order to keep the algorithm simple, we will assume that this term can be split as $\avg{b_1(\nu,\beta)\mu_1(\nu,\beta)|\nu} = \avg{b_1(\nu,\beta)|\nu}\avg{\mu_1(\nu,\beta)|\nu} = b_1^{\rm (msd)}\avg{\mu_1|\nu}$. This assumption can be explicitly tested using $N$-body simulations. With this caveat, the reconstructed quadratic coefficient is
\begin{align}
\delc^2 b_{20}^{\rm (msd)}(\nu) &= \frac1{(1-\epc)^2}\bigg[\frac{\delc^2 b_2^{\rm (msd)}}{(\Sc/\So)^2}\notag\\ 
&\ph{\bigg[\mu_2\bigg]}
- 2\epc\left(\frac{\delc b_1^{\rm (msd)}}{(\Sc/\So)}\avg{\mu_1|\nu} - \avg{\mu_1^2|\nu}\right)\notag\\
&\ph{(1-\epc)^2\bigg[2\epc\mu_1\lam_1\bigg]}
- \epc(2-\epc)\avg{\mu_2|\nu} \bigg]\,.
\label{dc2b20-avg}
\end{align}
The remaining averages over $\beta$ simplify to some extent, and we have
\begin{align}
\avg{\mu_1|\nu} &= \nu\left(\nu+\avg{\beta|\nu}\right)\,,\notag\\
\avg{\mu_2|\nu} &= \avg{\mu_1^2|\nu} - \nu^2\notag\\
&=\nu^2\left(\nu^2-1-2\nu\avg{\beta|\nu}+\avg{\beta^2|\nu}\right)\,,
\label{mu1mu2-avg}
\end{align}
where
\be
\avg{\beta^j|\nu} = \int\der\beta\,p(\beta|\nu)\,\beta^j\,,
\label{beta-avg}
\ee
with $p(\beta|\nu)$ given in \eqn{pbetanu}.

\section{Discussion}
We have presented a derivation of equation~(\ref{fESP}), which represents the first analytic model of halo abundances which accounts self-consistently for the fact that the physics of collapse is best described around the halo center of mass, is not spherical, and can vary substantially from one place in the universe to another.  The comparison between \eqn{fESP} and halo counts in simulations (Figure~\ref{ratio}) is very encouraging.  (Figure~\ref{vfv-const} shows that accounting for the physics of non-spherical collapse is necessary.)  

We also derived expressions for the associated large scale bias factors $b_{10}$ and $b_{20}$ (these follow from setting $\epc\to0$ and $\Sc/\So\to1$ in equations~\ref{dcb1} and~\ref{dc2b2}); the former is in very good agreement with measurements from simulations.  We are in the process of determining if $b_{20}$ is similarly accurate. Additionally, we showed how real-space measurements of $b_1$ and $b_2$ can be converted to practical estimates of $b_{10}$ and $b_{20}$ (equations~\ref{dcb10-avg} and~\ref{dc2b20-avg}), which extends the results of \citet{mps12} and \citet{ps12} to the case of stochastic barrier heights. Testing the accuracy of this reconstruction in $N$-body simulations is also work in progress.

Our ESP analysis highlighted the benefits of developing a model for the abundance of positions in space around which the physics of collapse is simplest.  In Section~\ref{allwalks}, we showed that it can be used to understand what would have happened if one had instead studied all initial positions (Figures~\ref{pdeltaPk} and~\ref{pdeltaPk2}).  This provides the basis for a number of other tests of the more formal aspects of our approach -- as well as elucidating the relation between our ESP calculation and the usual excursion set approach.  

\begin{figure}
 \centering
 \includegraphics[width=\hsize]{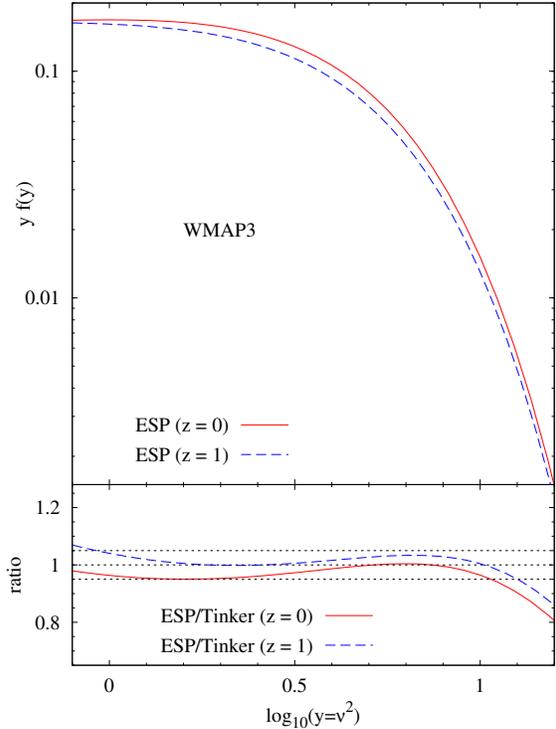}
 \caption{Predicted departures from self-similarity in the ESP model, for the WMAP3 cosmology.  
          Top panel shows the predicted halo abundances at $z=0$ and $1$, 
          expressed as a function of $y= [\delta_c D(0)/D(z)\sigma_0(m)]^2$. 
          We have assumed that the distribution $p(\beta)$ defined in the text does not evolve
          with redshift.
          Bottom panel shows the ratio of our predictions to the fitting 
          formulae for these redshifts which \citet{t+08} calibrated 
          from halo abundances measured in their simulations.}
 \label{vfvz}
\end{figure}

Our model must take as input one physical quantity from simulations.   This quantity describes how the typical density required for collapse depends on proto-halo mass, and how much this density varies from one proto-halo to the next.  While the triaxial collapse model correctly predicts that this mean value and the scatter around it should increase as halo mass decreases, the predicted scatter is smaller than observed.  We therefore hope that our work will motivate studies of the physical origin of $p(\beta)$, as well as measurements in simulations of how this distribution evolves.  If this distribution does not evolve, then our ESP model provides a particularly simple model for how the distribution of halo abundances, when expressed as a function of $\nu$, should be nearly but not exactly universal \citep[see discussion in][]{ms12}.  Figure~\ref{vfvz} shows that the predicted departures from universality in this simple case are in reasonable agreement with measurements.  

Returning to the issue of the distribution of $\beta$, an obvious direction of future work is that peaks have non-trivial deformation \emph{and} inertia tensors. Non-spherical initial shapes mean the critical density for collapse will now depend not just on the tidal field but on the shape and the misalignment angle between the principal axes of the two tensors. In principle, our ESP model contains a prescription for the distribution of shapes (and misalignments) for which one should run the ellipsoidal collapse model.  

This does raise one caveat:  in peaks theory (on which our ESP model is based), the axes of the two tensors are expected to be well aligned, with the direction of maximum compression (due to the tidal field) aligned with the shortest axis \citep{vdwb96}.  Measurements in simulations do show strong alignment, but the direction of maximum compression lies instead along the longest axis \citep{pdh02, dts13}.  On the other hand, these measurements are dominated by masses for which $\delc(m)/\sig_0(m) < 1$, where our ESP model may not apply anyway:  there is no physically compelling reason why a peak of (normalized) height 1 should dominate its surroundings. 
While this mass regime may be less interesting for studies which use the abundance of rich clusters to constrain cosmological models, understanding it is useful for understanding the formation and evolution of galaxies.  So we hope our approach of identifying just what it is that is special about the positions around which gravitational collapse occurs, and then using only these positions to make inferences about halo assembly histories, will eventually provide insight into the low mass regime as well.

\section*{Acknowledgements}
We thank Christophe Pichon for discussions, and for pointing out to us the work of \citet{h01}, and Ixandra Achitouv for discussions regarding barrier stochasticity. This work is supported in part by NSF-0908241 and NASA NNX11A125G (RKS, AP), and by the Swiss National Science Foundation (VD). AP thanks A. R\'efr\'egier and A. Amara at ETH, Z\"urich for hospitality during a visit when this work was completed. VD also wishes to thank ICTP for hospitality when parts of this work were completed.

\label{lastpage}

\end{document}